\documentclass[aip,amsmath,amssymb,jcp,preprint]{revtex4-2}

\usepackage{color}
\usepackage{dcolumn}% Align table columns on the decimal point
\usepackage{graphicx}
\usepackage{tabularx}
\usepackage{enumerate}
\usepackage{dsfont}
\usepackage{mathrsfs}
\usepackage[dvipsnames]{xcolor}
\usepackage{mathtools}

\usepackage[T1]{fontenc}

\usepackage[hidelinks]{hyperref}

\usepackage{afterpage}

\newcommand{\adcf}{ADC(2,2)$_f$}
\newcommand{\adcm}{ADC(2,2)$_m$}

\newcommand{\Psub}{\mathcal{P}}

\newcommand{\Pspace}{\mathcal{P}}
\newcommand{\Qspace}{\mathcal{Q}}
\newcommand{\lsq}{\mathcal{L}^2}

  \newcommand{\ket}[1]{|#1 \rangle}
  \newcommand{\bra}[1]{\langle #1 |}

  \newcommand{\myint}[1]{\int\!\!d{#1}\,}

  \newcommand{\eV}{\mathrm{eV}}
  \newcommand{\meV}{\mathrm{meV}}

\newcommand{\isr}[1]{\tilde{\Psi}_{#1}^{(N-1)}}
\newcommand{\diisr}[1]{\tilde{\Phi}_{#1}^{(N-2)}}

\newcommand{\DIs}[1]{\varphi^{(N-2)}_{#1}}
\newcommand{\dadbr}{dAD\ BR} %{BR$_\mathrm{dAD}$}
\newcommand{\mdadbr}{\mathrm{BR}_\mathrm{dAD}}

\newcolumntype{L}[1]{>{\raggedright\arraybackslash}p{#1}}
\newcolumntype{C}[1]{>{\centering\arraybackslash}p{#1}}
\newcolumntype{R}[1]{>{\raggedleft\arraybackslash}p{#1}}

\begin{document}
\title{Ab initio calculations of molecular double Auger decay rates} %Title of paper

\author{P.\ Koloren\v{c}}
\email[]{premysl.kolorenc@matfyz.cuni.cz}
% \homepage[]{}
%\thanks{}
\affiliation{Charles University, Faculty of Mathematics and Physics, Institute of Theoretical Physics, V Hole\v{s}ovi\v{c}k\'ach 2, 180 00 Prague, Czech Republic}

\date{\today}

\begin{abstract}
We report on the application of the recently developed Fano-ADC(2,2) method to compute total and partial Auger decay widths of molecular core-hole states, including explicit evaluation of double Auger decay branching ratios. The method utilizes the fast-convergent intermediate state representation to construct many-electron wave functions and is readily applicable to atoms, molecules and clusters. The ADC(2,2) scheme describes the initial and final states of the normal Auger decay consistently up to the second order of perturbation theory. In addition, excitations with two electrons in the continuum provide access to three-electron decay modes. The method yields decay widths and the Auger electron spectra in excellent agreement with the experiment, demonstrating the high accuracy of partial widths. The average relative error of double Auger decay branching ratios compared to available experimental data is about 30\%, which should be evaluated as an excellent result considering the universality of the method, the complexity of the double decay process and the neglection of nuclear motion in the present study.
\end{abstract}

\pacs{}% insert suggested PACS numbers in braces on the following line

\maketitle %\maketitle must follow title, authors, abstract and \pacs

% {\color{red}\tiny
% \section{TODO}
% \begin{enumerate}
%      \item reanalyze the projectors
%     \item ADC(2,2)$_m$ calculations for all molecules for comparison
%     \item Can I afford CH$_3$F at some level? Is it available in the Roos paper?
%     \begin{itemize}
%         \item the molecule is not there, no reference number, then
%         \item but there is CF$_4$ for both C and F $1s$ hole; maybe I could try this one with the same idea
%     \end{itemize}
%     \item compute HCl
%     \item should I include some Auger spectrum? Which? Add (2,2)$_m$ variant as well
% \end{enumerate}
% }

% %######### WORK NOTES ##########
% \input{work}
% %######### WORK NOTES ##########

\section{Introduction}
\label{sec:Intro}

Inner-shell ionized or excited low-Z atoms and molecules relax predominantly by autoionization processes, such as Auger decay\cite{Meitner1922zp,Auger1925jpr} (AD). When the excited moiety is embedded in an environment, interatomic Coulombic decay\cite{Cederbaum1997prl,Jahnke2020chrev} offers an alternative pathway, which becomes particularly significant for inner-valence excitations when the local AD is energetically forbidden. At the most fundamental level, investigating these phenomena helps understand the correlation in bound systems. More practically, Auger electron spectroscopy is one of the most widely used analytical tools in many branches of applied physics. Both aforementioned processes also play an important role in the damage caused by radiation to organic matter. Indeed, a significant share of the damage is inflicted indirectly by reactive secondary particles produced by AD or similar mechanisms. 

When the initial excitation energy is above the triple or even higher ionization threshold, processes involving multi-electron transitions become possible. In double Auger decay\cite{Carlson1965prl} (dAD) or double ICD\cite{Averbukh2006prl} (dICD), two electrons are emitted following recombination of a single inner-shell vacancy. These higher-order processes are often of very low relative intensity. However, under favourable conditions, they can become significant or even dominant decay channels. Recent works by Roos \emph{et al}\cite{Roos2016pccp,Roos2018scirep} showed that, in molecules, triple ionization via dAD can account for as much as 25\% of the decay following the formation of a single core vacancy. dICD was observed to be a relevant relaxation mechanism in endohedral fullerenes\cite{Obaid2020prl} or alkali dimers embedded in helium droplets\cite{LaForge2019nphys}.

In atoms, the mechanism of dAD is well understood through many-body perturbation theory\cite{Amusia1992pra}. The two electrons can be emitted either sequentially in a cascade of two-electron transitions or simultaneously. In the latter case, the relevant mechanisms are shake-off and knock-out. In shake-off, after the ejection of the first Auger electron, the second electron can relax into a continuum state due to orbital relaxation in the perturbed system. In the knock-out mechanism (virtual inelastic scattering), the first Auger electron collides with another outer-shell electron, which is also ejected to the continuum. Most theoretical studies of atomic dAD rely on an independent evaluation of these three contributions \cite{Zeng2013pra,Ma2017pra}. See Ref.~\onlinecite{Kolorenc2016jpb} for a comprehensive review.

Investigation of dAD in molecules is scarce. Prior to the works of Roos and coworkers cited above, two-electron emission was observed after carbon $1s\rightarrow 2\pi^*$ excitation of CO\cite{Hitchcock1988pra,Journel2008pra}. Using three-electron coincidence spectroscopy, Eland and coworkers studied the decay of core-ionized states of methane, OCS, and CS$_2$ \cite{Eland2010chpl, Eland2010jcp14, Eland2010jcp104}, providing vibrationally resolved triple ionization spectra. Evidence for strong three or more electron detachment in the decay of a K-shell vacancy in gas-phase DNA was recently given by Li \emph{et al.}\cite{Li2021chsci}. To our knowledge, there is no theoretical study of molecular dAD rates except the statistical approach employed in Ref.\ \onlinecite{Eland2010jcp14}.

In the present work, we apply the recently developed Fano-ADC(2,2) method\cite{Kolorenc2020jcp} to compute molecular AD widths, AD electron spectra and dAD branching ratios. Building on the work of Howat \emph{et al.}~\cite{Howat1978jpb}, the Fano-ADC method was initially devised and first applied to study AD and ICD by Averbukh and Cederbaum in 2005 \cite{Averbukh2005jcp}. Over the years, the approach was further developed and generalized for a broader range of problems, namely ICD in doubly ionized \cite{Kolorenc2008jcp} as well as neutral excited clusters \cite{Gokhberg2006jcp, Kopelke2011jcpB}, or Penning ionization \cite{Yun2018jcp}. Generic implementation, first applied by Koloren\v{c} and Sisourat in Ref.~\onlinecite{Kolorenc2015jcp} to ICD in helium trimer, is suitable for systems with arbitrary symmetry. The most notable applications of the Fano-ADC method comprise the study of ICD in helium dimer \cite{Sisourat2010nphys}, electron transfer-mediated decay in NeKr$_2$ \cite{Stumpf2013prl}, resonant Auger-ICD cascades \cite{Gokhberg2014nat}, collective decay in fluoromethane \cite{Feifel2016prl} or core-ICD in rare gases \cite{Kustner2023CommPhys}.

The method is based on the formalism of the Fano theory of resonances \cite{Fano1961pr, Feshbach1964rmp}. It relies on the explicit construction of the bound-like discrete state representing the metastable state (resonance) coupled to continuum-like final states. The many-electron wave functions are constructed in terms of an $\lsq$ one-particle basis using size-consistent and fast-convergent algebraic diagrammatic construction (ADC) methodology \cite{Schirmer1982pra, Schirmer1983pra} in the intermediate states representation (ISR) \cite{Schirmer1991pra, Mertins1996pra}. Correct normalization and interpolation of the discretized continuum is achieved via the Stieltjes imaging technique \cite{Langhoff1979empmc, Hazi1979empmc}. The present application to the study of double decay rests on the ability of the ADC(2,2) scheme to represent the two electrons in the continuum explicitly.

% \enlargethispage{1cm}
The method is described in Ref.~\onlinecite{Kolorenc2020jcp}, so only a brief overview is given in Sec.~\ref{sec:FanoADC}. In~Sec.~\ref{sec:BRs}, we discuss the evaluation of channel-specific branching ratios (BR) using approximate channel projectors. These are constructed from dicationic states represented at the ADC(2)x level of theory\cite{Schirmer1984zpa}, consistently with the level of description of the final states in ADC(2,2). In Sec.~\ref{sec:Results}, we report on the application of the method to evaluate dAD BRs in neon-isoelectronic molecules and small hydrocarbons and provide an in-depth analysis of the results. The work is summarized in Sec.~\ref{sec:conclusions}. Atomic units $e=m_e=\hbar=1$ are used.

\section{Method}
\label{sec:method}

\subsection{Fano-ADC(2,2)}
\label{sec:FanoADC}

Within the Fano formalism, the exact continuum solution $\ket{\psi(E)}$ of the Schr\"odinger equation 
\begin{equation}
    \hat{H}\ket{\psi(E)}=E\ket{\psi(E)}
\end{equation}
in the spectral region around the energy of the resonance is described as a bound-like discrete state $|\phi_d\rangle$ embedded in $N_c$ continua $|\chi_{\beta}(\epsilon)\rangle$ associated with the individual decay channels,
\begin{equation}
    \label{eq:Fano_ansatz11}
    \ket{\psi(E)} = a(E)\ket{\phi_d}+\sum_{\beta=1}^{N_c}\myint{\epsilon} b_\beta(E,\epsilon)\ket{\chi_\beta(\epsilon)}.
\end{equation}
Specifically for AD, $|\phi_d\rangle$ represents the inner-shell ionized metastable state, and $|\chi_\beta(\epsilon)\rangle$ is the continuum wave function associated with the di- or tricationic decay channel $\beta$ with threshold energy $E_\beta$, and $\epsilon$ is the energy of the continuum electron.\footnote{For the tricationic decay channels, the integration over $\epsilon$ is to be replaced by a double integral over the energies of the two outgoing electrons.} In the rigorous theory, the total decay width is a function of energy given as a sum of \text{partial decay widths},
\begin{equation}
    \label{eq:gamma_E}
    \Gamma(E) = \sum_\beta \Gamma_\beta(E) = 2\pi\sum_\beta |\langle\phi_d|\hat{H}-E_d|\chi_\beta(E-E_\beta)\rangle|^2.
\end{equation}
The decay width associated with the Gamow-Siegert state \cite{Siegert1939pr} and related to the inner-shell vacancy lifetime can be approximated by the value of $\Gamma(E)$ at the discrete state energy $E_d$,
\begin{equation}
    \label{eq:gamma_loc}
    \Gamma_{loc} = \frac{1}{\tau}\approx \Gamma(E=E_d).
\end{equation}
In the following, we drop the subscript $loc$ and the distinction between the local decay width $\Gamma$ and the width function $\Gamma(E)$ will be indicated explicitly by energy dependence.

% \subsection{Algebraic diagrammatic construction\label{sec:ADC}}

To evaluate the formula \eqref{eq:gamma_E}, explicit representation of the initial and final multi-electron wave functions, $\ket{\phi_d}$ and $\ket{\chi_\beta(\epsilon)}$, has to be provided. Within the ISR-ADC framework, this is achieved via configuration interaction-like expansion in terms of the so-called \emph{intermediate states} (IS) $\ket{\isr{J}}$. ISs are derived by application of the physical excitation operators $\hat{C}_J$,
\begin{equation}
    \label{eq:C_J}
    \left\{\hat{C}_J^{(N-1)}\right\}=\left\{c_k;\ c^\dag_a c_k c_l,k<l;\right. \\
    \left. c^\dag_a c^\dag_b c_j c_k c_l,j<k<l, a<b;\dots \right\},
\end{equation}
to the perturbation-theoretically (PT) corrected ground state wave function $\ket{\Psi_0}$ of the neutral system, followed by \emph{excitation class orthogonalization} (ECO) procedure,\cite{SchirmerMBM2018}
\begin{equation}
    \label{eq:ISs}
    \ket{\Psi^0_J} = \hat{C}_J \ket{\Psi_0} \xrightarrow{\mathrm{ECO}}\ket{\isr{J}}.
\end{equation}
Here, $\ket{\Psi^0_J}$ are non-orthogonal so-called correlated excited states,
$c_q$ ($q=i,j,k\dots$) stand for the annihilation operators associated with \emph{occupied} (hole) HF spin-orbitals and $c^\dag_q$ ($q=a,b,\dots$) for the creation operators associated with \emph{virtual} (particle) HF spin-orbitals; $J$ is the corresponding multi-index. Specifically, in the ADC(2,2) scheme employed in the present work, the PT expansion of the ground state is truncated after 2$^{nd}$ order,
\begin{equation}
    \label{eq:GS_PT}
    \ket{\Psi_0}=\ket{\Phi_0}+\ket{\Psi_0^{(1)}}+\ket{\Psi_0^{(2)}},
\end{equation}
and the physical excitation operators are restricted to the lowest three excitation classes -- $1h$ (one-hole), $2h1p$ (two-hole-one-particle) and $3h2p$, given explicitly in Eq.~\eqref{eq:C_J}.
%Compared to the configuration interaction method, where the basis configuration functions are derived from the uncorrelated single-determinant Hartree-Fock (HF) ground state $\ket{\Phi_0}$, the ISR-ADC approach leads to faster convergence of the cationic wave function expansion.
% \begin{equation}
%     \label{eq:C_J}
%     \left\{\hat{C}_J^{(N-1)}\right\}=\left\{c_k;\ c^\dag_a c_k c_l,k<l;\ c^\dag_a c^\dag_b c_j c_k c_l,j<k<l, a<b \right\}.
% \end{equation} 
% Here, $c_q$ ($q=i,j,k\dots$) stand for the annihilation operators associated with \emph{occupied} (hole) HF spin-orbitals and $c^\dag_q$ ($q=a,b,\dots$) for the creation operators associated with \emph{virtual} (particle) HF spin-orbitals; $J$ is the corresponding multi-index.

The definition \eqref{eq:ISs} together with the expansion of the reference wave function \eqref{eq:GS_PT} results naturally in PT expansions of the ISs, which are truncated after the second order for the $1h$ and $2h1p$ ISs and after the zero order for $3h2p$ ISs. In turn, the elements of the matrix representation of the Hamiltonian shifted by the ground state energy $E_0$,
\begin{equation}
    \label{eq:MIJ_exact}
    M_{IJ}=\bra{\isr{I}}\hat{H}-E_0\ket{\isr{J}},
\end{equation}
are given in terms of a perturbation series and can be tailored to provide ionization energies as its eigenvalues consistently through the desired order of PT. In particular, the \adcf\ and \adcm\ schemes (cf Tab.\ I in Ref.\ \onlinecite{Kolorenc2020jcp}) are designed such that the ionization energies of $1h$- and $2h1p$-like states (i.e., the initial inner-shell hole and the final states of the normal AD) are evaluated up to the second order of PT. The energies of the $3h2p$-like states (i.e., the final states of dAD) are computed only through the first and zeroth order in \adcf\ and \adcm, respectively.

To provide approximations for the bound-like $|\phi_d\rangle$ and continuum-like $\ket{\chi_\beta(\epsilon)}$ components required for the evaluation of the decay width function Eq.\ \eqref{eq:gamma_E}, the configuration space spanned by ISs is divided into the subspace $\Pspace$ containing configurations associated with open decay channels and the complementary subspace $\Qspace$. To this end, we employ the universal procedure referred to as \emph{selection scheme B}
%\footnote{The number of open dicationic and tricationc channels is determined by ADC(2)x calculations for double and triple ionization, respectively.}
in Ref.\ \onlinecite{Kolorenc2020jcp}. The initial inner-shell hole state is then represented by a selected eigenvector of the Hamiltonian matrix projected onto the $\Qspace$ subspace, $\mathbf{QMQ}$. The final states of the decay are represented by the eigenvectors $|\chi_i\rangle$ with eigenvalues $\epsilon_i$ of the $\mathbf{PMP}$ Hamiltonian matrix. $\mathbf{M}$, $\mathbf{Q}$ and $\mathbf{P}$ are ISR matrices of the shifted Hamiltonian \eqref{eq:MIJ_exact} and the projectors onto the $\Qspace$ and $\Pspace$ subspaces, respectively.

Since a square-integrable Gaussian single-particle (GTO) basis is used, the decay continuum is discretized, i.e., the resolution of identity in the $\Psub$ subspace reads
\begin{equation}
    \label{eq:Presolution}
    \sum_{\beta=1}^{N_c}\myint{\epsilon}\ket{\chi_\beta(\epsilon)}\bra{\chi_\beta(\epsilon)}\approx \sum_i\ket{\chi_i}\bra{\chi_i},\  \langle\chi_i|\chi_j\rangle = \delta_{ij}.
\end{equation}
In turn, the decay width function \eqref{eq:gamma_E} cannot be evaluated directly. Instead, we evaluate discrete couplings
\begin{equation}
    \label{eq:gamma_i}
    \gamma_i = 2\pi |\bra{\phi_d}\mathbf{M}-E_d\ket{\chi_i}|^2
\end{equation}
and use the relation \eqref{eq:Presolution} to compute inverse spectral moments of the decay width function,
\begin{equation}
    \label{eq:spectral}
    S_{-k}=\int\! E^{-k}\Gamma(E)dE \approx \sum_i (\epsilon_i)^{-k} \gamma_i.
\end{equation}
Stieltjes imaging technique \cite{Langhoff1979empmc, Hazi1979empmc, Reinhardt1979cpcom} is employed to recover $\Gamma(E)$ from the spectral moments~\eqref{eq:spectral}. Example outputs of the Stieltjes imaging procedure can be found in  Supplementary Material (SM).

\subsection{Branching ratios}
\label{sec:BRs}

The Eq.\ \eqref{eq:Presolution} indicates that the discrete eigenvectors $\ket{\chi_i}$ cannot be assigned to individual decay channels, and the breakdown of the total decay width function into partial widths $\Gamma_\beta$ apparent in Eq.\ \eqref{eq:gamma_E} is lost. Partial decay widths can still be estimated by introducing approximate channel projectors $\hat{P}_\beta^{(N-2)}$ and repeating the Stieltjes imaging technique for each channel individually using projected eigenvectors $\hat{P}_\beta^{(N-2)}\ket{\chi_i}$ and partial couplings,\cite{Averbukh2005jcp, Stumpf2017chp}
\begin{equation}
    \label{eq:gamma_i_partial}
    \gamma_i^{(\beta)}=2\pi|\bra{\phi_d}\mathbf{M P}_\beta\ket{\chi_i}|^2.
\end{equation}
This procedure yields unnormalized estimates of the partial decay widths, $\Tilde{\Gamma}_\beta=\tilde{\Gamma}_\beta(E=E_d)$. Finally, the partial widths are normalized to the total width $\Gamma = \Gamma(E=E_d)$ obtained from full couplings \eqref{eq:gamma_i},
\begin{equation}
    \label{eq:Gamma_beta_normalized}
    \Gamma_\beta = \Gamma\frac{\tilde{\Gamma}_\beta}{\sum_\beta \tilde{\Gamma}_\beta}.
\end{equation}

For two-electron AD, the decay channels are defined by dicationic states $\ket{\DIs{\beta}}$, which can be either of the $2h$ character for main channels or $3h1p$-like for shake-up satellites. Within the Fano-ADC(2,2) framework, corresponding final states $\ket{\chi_i}$ are dominated by $2h1p$ and $3h2p$ ISs, respectively. Consistently with the ADC(2,2) level of representation of the final states, the dicationic states $\ket{\DIs{\beta}}$ can be expanded in terms of $2h$ and $3h1p$ ISs $\ket{\tilde{\Phi}_I^{(N-2)}}$ derived from the second-order ground state \eqref{eq:GS_PT} by application of the excitation operators
\begin{equation}
    \label{eq:C_Jdicationic}
    \left\{\hat{C}_I^{(N-2)}\right\}=\left\{c_k c_l,k<l;\ c^\dag_a c_j c_k c_l,j<k<l\right\}
\end{equation}
followed by the ECO procedure. The resulting PT expansions are truncated after the second order for the $2h$ ISs and after the zero order for the $3h1p$ ISs. This corresponds exactly to the extended second-order scheme for double ionization  [DI-ADC(2)x].
%\footnote{To be strictly consistent, the DI-ADC(2)x scheme is appropriate for channels definition in the Fano-\adcf\ method while the strict DI-ADC(2) scheme should be combined with Fano-\adcm.}
In Ref.~\onlinecite{Schirmer1984zpa}, the \mbox{DI-ADC(2)x} is derived using the diagrammatic technique rather than within the ISR framework. However, the formulas given for the Hamiltonian matrix elements hold for both formalisms. For more details on the relation between the diagrammatic and ISR formulations of ADC, see Ref.~\onlinecite{SchirmerMBM2018}.

By diagonalization of the ISR-ADC(2)x Hamiltonian matrix for the $(N-2)$-electron system, we obtain the channel energies $E_\beta^{(N-2)}$ and the ISR expansions of the corresponding dicationic states,
\begin{equation}
    \label{eq:channel_functions}
    \ket{\DIs{\beta}} = \sum_I q_\beta^I \ket{\diisr{_I}}.
\end{equation}
The channel projectors are defined as
\begin{equation}
    \label{eq:P_beta_general}
    \hat{\tilde{P}}_\beta^{(N-2)}=\sum_\epsilon c^\dag_\epsilon \ket{\DIs{\beta}}\bra{\DIs{\beta}}c_\epsilon \\ = \sum_\epsilon\sum_{I,I'}q^I_\beta (q^{I'}_\beta)^* c^\dag_\epsilon \ket{\diisr{I}}\bra{\diisr{I'}}c_\epsilon.
\end{equation}
In the above expression, $I$ is a multi-index running over the $2h$ and $3h1p$ ISs, and $c^\dag_\epsilon$ is a creation operator associated with a continuum-like spin-orbital. In the present implementation, which is based on purely Gaussian-type (GTO) basis sets, there is no distinction between a continuum and bound virtual spin-orbitals. Therefore, the complete set of virtual spin orbitals, $c_\epsilon^\dag\in\{c^\dag_a\}$, is used in the projector definition.

As a result of this choice, the vectors $c^\dag_a \ket{\diisr{I}}$ are not perfectly orthogonal for $a\neq a'$ or $I\neq I'$, or both. This non-orthogonality has the same origin as the non-orthogonality of the correlated excited states $\hat{C}_J\ket{\Psi_0}$ [cf.~Eq.~\eqref{eq:ISs}] and stems from the appearance of the virtual orbitals in higher-order contributions to the correlated reference ground state $\ket{\Psi_0}$~\eqref{eq:GS_PT}. Consequently, the projectors $\hat{\tilde{P}}_\beta^{(N-2)}$ defined by Eq.~\eqref{eq:P_beta_general} are not exact projectors, $\left(\hat{\tilde{P}}_\beta^{(N-2)}\right)^2\neq \hat{\tilde{P}}_\beta^{(N-2)}$, and neither are they orthogonal for different channels.
%, $\hat{\tilde{P}}_\beta^{(N-2)}\hat{\tilde{P}}_{\beta'}^{(N-2)}\neq \delta_{\beta\beta'}\hat{\tilde{P}}_\beta^{(N-2)}$.
It is easy to show that the ECO-like procedure applied to the $c^\dag_a \ket{\diisr{I}}$ vectors would transform the set into the $(N-1)$-electron ISs forming the basis of ADC(2,2) configuration space,
\begin{equation}
    \label{eq:ECO_on_epsDIIS}
    \left\{c_a^\dag\ket{\diisr{I}}\right\}\xrightarrow{\mathrm{ECO}}\left\{\ket{\isr{J}}\right\}
\end{equation}
with the straightforward relation between the $I$ and $J$ multi-indices, e.g., $I=kl \rightarrow J=akl$, or $J=aI$. It is thus possible to define a set of orthogonal proper projectors as
\begin{equation}
    \label{eq:P_beta_orthog}
    \hat{P}_\beta^{(N-2)}=\sum_a\sum_{I,I'} q^I_\beta (q^{I'}_\beta)^* \ket{\isr{J=aI}}\bra{\isr{J'=aI'}}.
\end{equation}
This approach, however, is not entirely correct either as $\hat{P}_\beta^{(N-2)}$ does not correspond precisely to the original dicationic state \eqref{eq:channel_functions}. On the other hand, it is computationally significantly cheaper as it avoids evaluation of $\bra{\diisr{I}}c_\epsilon\ket{\isr{J}}$ matrices and simplifies the construction of the complementary projector $\hat{P}_\perp$ defined below, which is needed for the evaluation of the dAD branching ratio. Therefore, we adopt this definition in the present work.

Similarly, the decay channels corresponding to dAD can be defined using tricationic states $\ket{\varphi_\eta^{(N-3)}}$ constructed in the framework of the first-order TI-ADC(1) scheme for triple ionization. It is equivalent to configuration interaction in terms of $3h$ HF configuration state functions,
\begin{equation}
    \label{eq:3h_configs}
    \ket{\Phi_{klm}^{(N-3)}}=c_k c_l c_m \ket{\Phi_0},\ k<l<m.
\end{equation}
The resulting projectors
\begin{equation}
    \label{eq:P_eta_TI}
    \hat{P}_\eta^{(N-3)}=\sum_{\epsilon<\epsilon'} c^\dag_\epsilon c^\dag_{\epsilon'}\ket{\varphi_\eta^{(N-3)}}\bra{\varphi_\eta^{(N-3)}}c_{\epsilon'}c_\epsilon
\end{equation}
are free from the non-orthogonality issues owing to the simplicity of the configurations \eqref{eq:3h_configs}.
However, to cover the full normal and shake-up AD spectrum, and to account properly for the sequential dAD, dicationic channels \eqref{eq:channel_functions} with energies $E_\beta^{(N-2)}$ well above triple-ionization potential (TIP) often have to be taken into account when building the projectors $\hat{P}_\beta^{(N-2)}$. Without the distinction between bound and continuum virtual orbitals, many such DI-ADC(2)x states $\ket{\varphi_\beta^{(N-2)}}$ with dominant $3h1p$ character will represent continuum states associated with some tricationic channel $\ket{\varphi_\eta^{(N-3)}}$. And vice versa, projectors \eqref{eq:P_eta_TI} will also contain contributions from shake-up states already included among dicationic channels.

As this ambiguity cannot be satisfactorily resolved within the present implementation of the Fano-ADC(2,2) method, we do not explicitly build projectors onto the tricationic channels. Rather, the dAD branching ratio (\dadbr) is determined by the following procedure:
\begin{enumerate}
    \item All DI-ADC(2)x eigenstates with eigenenergies up to some energy $E^{(N-2)}_\mathrm{max}>\mathrm{TIP}$ are used to construct channel projectors \eqref{eq:P_beta_orthog}. The maximum energy
    %must be above the triple-ionization threshold and
    is usually determined by the highest shake-up dicationic state with $2h$ strength of at least 1\%.
    \item A complementary projector $\hat{P}_\perp=\hat{P}-\sum_\beta \hat{P}_\beta^{(N-2)}$ is constructed, where $\hat{P}$ is the projector onto the full $\Pspace$ subspace. As the $\Pspace$ subspace contains only open channels and all dicationic decay channels up to $E_\mathrm{TIP}$ are included in step 1, this projector covers the missing contribution to the double decay.
    \item Corresponding partial decay widths $\Gamma_\beta$ and $\Gamma_\perp$ are calculated using Eqs.\ \eqref{eq:gamma_i_partial} and \eqref{eq:Gamma_beta_normalized}.
    \item the total \dadbr\ is determined from the sum of the partial decay widths associated with channels above TIP plus the complementary $\Gamma_\perp$,
    \begin{equation}
        \label{eq:dAD BR}
        \mathrm{BR}_\mathrm{dAD}=\frac{\Gamma_\perp+\sum_{E_\beta>\mathrm{TIP}}\Gamma_\beta}{\Gamma}.
    \end{equation}
\end{enumerate}

Using this procedure, it is implicitly assumed that any dicationic state above TIP populated by AD decays further by electron emission and contributes to cascade dAD. Any possible radiative decay or production of higher-charge ions, as well as quenching of the cascade decay by nuclear dynamics, is neglected. The approach also cannot distinguish between the simultaneous and cascade dAD. Thus, we only report total dAD branching ratios in the present work. This shortcoming can only be resolved by classifying the ISR-ADC correlated states for all the relevant charge states as bound or continuum. To this end, new tools would have to be developed, but such classification will likely be indistinct within a GTO basis. The problem could potentially be resolved by the restricted correlation space ADC approach briefly discussed below. 

\subsubsection*{Restricted Correlation Space ADC}
The orthogonality problem in the definition of the channel projectors Eq.~\eqref{eq:P_beta_general} can be circumvented in a \emph{restricted correlation space} ADC (RCS-ADC) approach introduced by M.\ Ruberti.\cite{Ruberti2019jctc} The space of one-particle virtual spin-orbitals is divided into two mutually orthogonal subspaces, $\{\hat{c}_a^\dag\}=\{\hat{c}^\dag_\alpha\}\oplus\{\hat{c}^\dag_\mu\}$. The bound-like virtual orbitals $\{\hat{c}^\dag_\alpha\}$ -- the RCS subspace -- together with the occupied orbitals $\{\hat{c}^\dag_i\}$ form a canonical set of molecular HF spin-orbitals. The second set, the \emph{ionization subspace} $\{\hat{c}^\dag_\mu\}$, comprise continuum-like virtual spin-orbitals, typically constructed using a B-spline basis set. The whole virtual orbital subspace is not necessarily canonical, i.e., the Fock operator is not diagonal and contains coupling between the RCS and ionization subspaces.

In the RCS-ADC formulation, the PT expansions of the reference ground state $\ket{\Psi_0}$ and the $(N-2)$-electron configuration space spanned by $\ket{\diisr{I_{RCS}}}$ are restricted to the RCS subspace. The $(N-1)$-electron $2h1p$ ISs $\ket{\isr{J}}$ can then be classified by the character of the virtual orbital as bound-like or continuum-like. Specifically for the continuum-like ISs with a virtual orbital from the ionization subspace $\mu$, it is easy to show  (c.f.\  Ref.~\onlinecite{Ruberti2019jctc}) that they are in the form of so-called augmented states,
\begin{equation}
    \label{eq:augmented states}
    \ket{\isr{\mu I}}=c^\dag_\mu\ket{\diisr{I}}.
\end{equation}
In turn, if the sum over virtual orbitals in the projector definition \eqref{eq:P_beta_general} can be restricted to the ionization subspace, $\sum_\epsilon \rightarrow \sum_\mu$, the resulting channel projectors will be naturally orthogonal exact projectors for an orthonormal set of dicationic states $\ket{\varphi_\beta^{(N-2)}}$. We will explore the RCS formulation in a future publication.

% {\color{red}
% Our preliminary calculations indicate that restricting the whole $\Psub$ subspace to the augmented states $\ket{\isr{\mu I}}$ results in severely underestimated total decay widths. If the RCS virtual orbitals $\hat{c}^\dag_\alpha$ play a significant role in the correlated continuum-like final states $\ket{\chi_i}$ as suggested by this observation, then the restriction of the channel projectors to the ionization subspace will be insufficient as well. The cause of this behaviour can probably be traced to the fact that the HF spin-orbitals are defined with respect to the neutral system and are inadequate for both the ionic and the dicationic states. In turn, the RCS virtual subspace needs to be relatively large for a satisfactory representation of the dicationic states $\ket{\varphi_\beta^{(N-2)}}$ and cannot be excluded from the expansion of the continuum-like cationic states $\ket{\chi_i}$ without compromising their quality.
% }

\section{Results and discussion}
\label{sec:Results}

We have used the \emph{full} Fano-\adcf\ and \emph{minimal} Fano-\adcm\ schemes to compute total AD decay widths and dAD branching ratios for a number of small core-ionized molecules. The main difference between \adcf\ and \adcm\ schemes is that the coupling between $3h2p$ ISs is neglected in the latter. This reduces the scaling of the computational cost from $n_{occ}^3 n_{virt}^4$ to $n_{occ}^4 n_{virt}^3$ ($n_{occ}$ and $n_{virt}$ being the number of active occupied and virtual molecular orbitals, respectively)\cite{Kolorenc2020jcp}, reducing the number of nonzero elements in the Hamiltonian matrix by \mbox{90\%--95\%}.

The choice of the systems was motivated primarily by the availability of experimental values in Ref.~\onlinecite{Roos2018scirep} and by computational manageability. In the present work, we want to explore the fundamental accuracy of the method. Therefore, we have used large basis sets to minimize basis set-related errors. The basis sets were optimized by obtaining the best possible stability of the Stieltjes imaging procedure and convergence of the decay width function $\Gamma(E)$ %and partial decay widths
using the computationally cheap Fano-ADC(2)x method. At the ADC(2,2) level, certain restrictions had to be imposed in many cases to reduce the size of the $3h2p$ configuration space. The restrictions have been carefully tested on small molecules and proved to have a lesser impact than using inferior one-particle basis sets. All the basis sets and configuration space-related computational details can be found in SM.

\subsection{Molecules isoelectronic to Ne}
\label{sec:isoNe}
For the initial benchmark, we have chosen the first row hydrides isoelectronic with Ne. The calculated total decay widths of the $1s^{-1}$ core hole states are summarised and compared to available experimental data in Tab.\ \ref{tab:isoNe_Gamma}. The errors given for calculated widths are estimated as the standard deviation of $\Gamma(E_d)$ obtained by the Stieltjes imaging procedure at different orders. %, see Figs.\ \ref{fig:SI.CH4.large} and \ref{fig:SI.CH4.small} in SM.
Thus, the error primarily reflects the stability of the Stieltjes imaging and the quality of the basis set, not any systematical errors inherent in Fano-ADC.

\begin{figure}
    \centering
    \includegraphics[width=\linewidth]{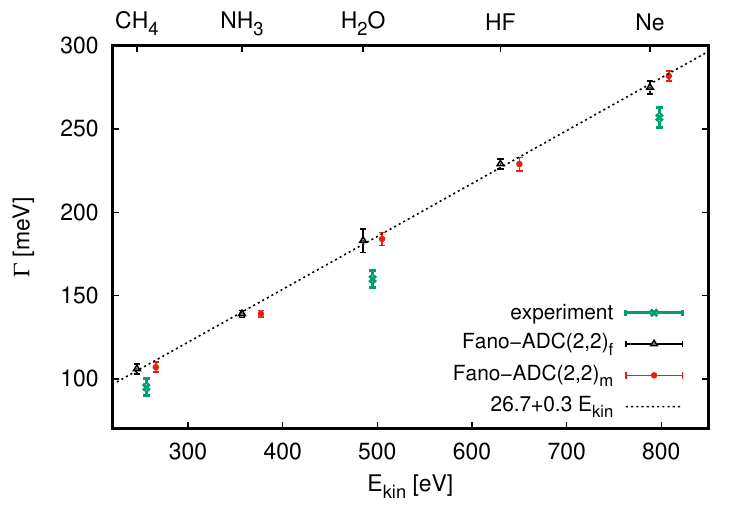}
    \caption{Total AD decay widths in Ne-isoelectronic molecules, plotted against the mean kinetic energy $\bar{E}_{kin}$ of the emitted electrons.}
    \label{fig:isoNe_Gamma}
\end{figure}

We observe that, on average, the more expensive Fano-\adcf\ scheme overestimates the experimental values of total AD widths by 10\%. The relative error is smallest for Ne, which indicates that the inaccuracy might be partially attributed to the missing effects of vibrational motion. However, the data clearly reveal a systematical bias of the method. The cheaper Fano-\adcm\ yields essentially equivalent results.
%The somewhat larger errors indicate a lower quality of the representation of the continuum subspace.
In Fig.\ \ref{fig:isoNe_Gamma}, we plot the same data against the mean kinetic energy of the emitted electrons,
\begin{equation}
    \label{eq:mean_e_kin}
    \bar{E}_{kin}=\frac{\myint{E}E\, I_{AD}(E)}{\myint{E} I_{AD}(E)},
\end{equation}
where $I_{AD}(E)$ is the calculated Auger electron spectrum. The graph shows a distinct linear trend for both the theoretical and experimental data.

\begin{figure}
    \centering
    \includegraphics[width=\linewidth]{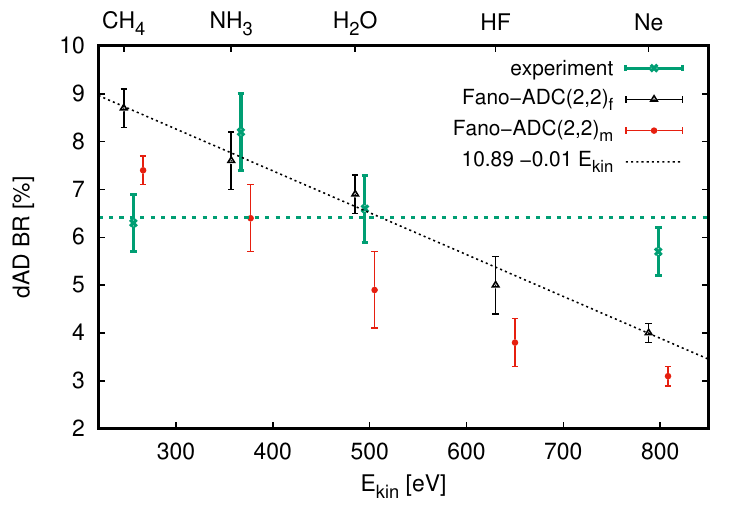}
    \caption{dAD branching ratios for Ne-isoelectronic molecules, plotted against the mean kinetic energy $\bar{E}_{kin}$ of the emitted electrons.}
    \label{fig:isoNe_dAD}
\end{figure}

Calculated and experimental values of \dadbr\ are compared in Tab.\ \ref{tab:isoNe_dAD} and Fig.\ \ref{fig:isoNe_dAD}.
The errors in the calculated branching ratios are determined from the uncertainty in the TIP, which enters the formula \eqref{eq:dAD BR}. The dAD process involves both $2h1p$-like and $3h2p$-like final states above TIP. However, their energies are evaluated inconsistently in \adcf\ -- through the second- and first-order PT, respectively. It is thus unclear what the appropriate value of TIP is. To reflect this uncertainty, we use both first- and second-order values of the TIP to determine the range reported in Tab.\ \ref{tab:isoNe_dAD}.

Focusing on the Fano-\adcf\ scheme, the relative discrepancy between computed and measured dAD BR ranges from 5\% to 38\%, averaging at 20\%, which should be classified as a very good result considering the complexity of the process and universality of the method. However, Fig.\ \ref{fig:isoNe_dAD} exposes a qualitative disagreement between the two sets of data. As for the total AD widths, the computed \dadbr\ depend linearly on the mean kinetic energy $\bar{E}_{kin}$. Experimental data do not really support such a trend. With the exception of NH$_3$, they remain roughly constant, in accord with the observation by Roos \emph{et al.}\cite{Roos2016pccp,Roos2018scirep} that \dadbr\ can be estimated from the number of available valence electrons $N_{ve}$ as
\begin{equation}
    \label{eq:dADBR_Roos}
    \mdadbr=A N_{ve}+B.
\end{equation}
This trend stems from the simple fact that the number of decay channels grows with the number of valence electrons approximately like $N_{ve}^2$ and $N_{ve}^3$ for normal and double AD, respectively. Using different samples of molecules to determine the coefficients $A$ and $B$, the formula predicts the values of \dadbr\  7.2\%\cite{Roos2016pccp} or 8.8\%\cite{Roos2018scirep}  for $N_{ve}=8$. Both values overestimate the average $6.4\%$ of the experimental dAD BRs listed in Tab.\ \ref{tab:isoNe_dAD}.

Moving to the cheaper Fano-\adcm\ scheme, we see from the table that it systematically recovers only about 80\% of the \dadbr\ computed by the full scheme. One of the reasons might be a further shift in the effective TIP. However, the consequences of neglecting the $3h2p/3h2p$ couplings are certainly more intricate as it disregards any correlation or interchannel coupling in the dAD final states, as well as the interaction between the emitted electrons.
%{\color{red} We will come back to this later, but deeper analysis shows that indeed the source of the errors is in the direct $1h-3h2p$ transitions; the stronger the mixing of $2h1p$-like and $3h2p$-like states above $E_\mathrm{TIP}$, the better \adcm\ works.}

%
\begin{table}[ht]
\renewcommand{\arraystretch}{1.3}
 \begin{tabular}{l|c||c|c|c}
 & $\bar{E}_{kin} (\eV)$ & Experiment & \adcf & \adcm \\
   \hline
   \hline
   Ne & 788 & $257\pm 6$\cite{Mueller2017aj} & $275\pm 4$ & $282\pm 3$ \\
   HF & 630 &                                & $229\pm 3$ & $229\pm 4$ \\
   H$_2$O & 485 & $160\pm 5$\cite{Sankari2003chpl} & $183\pm 7$ & $184\pm 4$ \\
   NH$_3$ & 357 &                            & $139\pm 2$ & $139\pm 2$ \\
   CH$_4$ & 246 & $95\pm 5$\cite{Carroll2002jcp} & $106\pm 3$ & $107\pm 3$ \\
   \hline
 \end{tabular}
 \caption{\label{tab:isoNe_Gamma} Total $1s^{-1}$ AD widths $\Gamma$ for the series of molecules isoelectronic to Ne. The second column shows the characteristic kinetic energy of secondary electrons, given by the first moment of the respective Auger electron spectrum. The third column lists available experimental values for $\Gamma$, fourth and fifth columns the widths computed using Fano-\adcf\ and Fano-\adcm, respectively. All widths are given in~meV. \\
 %{\color{red}There is 2.2fs (+0.2/-0.3) = 299meV result for Ne in https://doi.org/10.1038/s41567-020-01111-0 - NaturePhys 2021 streaking exp \\
 % some theoretical widths: Inhester H2O paper: Ne 269\,meV, H2O 163\,meV
 % }
}
\end{table}
\begin{table}[ht]
\renewcommand{\arraystretch}{1.3}
 \begin{tabular}{l|c||c|c|c}
 % \multicolumn{2}{c}{} & \multicolumn{3}{c}{dAD branching ratio (\%)} \\
 % \hline
 & $\bar{E}_{kin} (\eV)$ & Experiment & \adcf & \adcm \\
   \hline
   \hline
   Ne & 788 & $5.7\pm 0.5$ & $4.0\pm 0.2$ & $3.1\pm 0.2$ (77.5\%)\\
   HF & 630 &                                & $5.0\pm 0.6$ & $3.8\pm 0.5$ (76.0\%) \\
   H$_2$O & 485 & $6.6\pm 0.7$ & $6.9\pm 0.4$ & $4.9\pm 0.8$  (71.0\%) \\
   NH$_3$ & 357 & $8.2\pm 0.8$ & $7.6\pm 0.6$ & $6.4\pm 0.7$  (84.2\%) \\
   CH$_4$ & 246 & $6.3\pm 0.6$ & $8.7\pm 0.4$ & $7.4\pm 0.3$  (85.1\%) \\
   \hline
 \end{tabular}
 \caption{\label{tab:isoNe_dAD} dAD branching ratios for the Ne-isoelectronic series of molecules. The second column shows the characteristic kinetic energy of secondary electrons, third to fifth columns \dadbr\ in \%. For the Fano-\adcm\ method, the number in parentheses gives the percentage of the branching ratio recovered compared to the Fano-\adcf\ value. All experimental values are taken from Ref.~\onlinecite{Roos2018scirep}. For Ne, a more recent value is cited from Ref.~\onlinecite{Roos2019pccp}.
}
\end{table}

%%%%%%%%%%%%%%%%%%%%%%%%%%%%%%%%%%%%%%%%%%%%%%%%%%%%%%%%

\subsection{Small hydrocarbons}
\label{sec:hydrocarb}
\begin{table*}[ht]
 \renewcommand{\arraystretch}{1.3}
 \begin{tabular}{l|c|c||c|c||c|c}
 % \multicolumn{2}{c}{} & \multicolumn{3}{c}{dAD branching ratio (\%)} \\
 % \hline
 & \multicolumn{2}{c||}{Experiment} & \multicolumn{2}{c||}{Fano-\adcf} & \multicolumn{2}{c}{Fano-\adcm} \\
 & $\Gamma (\meV)$ & \dadbr (\%) & $\Gamma (\meV)$ & \dadbr (\%) & $\Gamma (\meV)$ & \dadbr (\%) \\
   \hline
   \hline
   CH$_4$ & $95\pm 5$\cite{Carroll2002jcp} & $6.3\pm 0.6$ & $107\pm 6$ & $8.2\pm 0.7$ & $109\pm 7$ & $7.8\pm 0.7$ \\ % (39\%)\\ 
   % \hline
   C$_2$H$_2$ & $106\pm 5$\cite{Carroll2002jcp} & $8.2\pm 0.8$ & $113\pm 8$ & $13.3\pm 1.4$ & $110\pm 8$ & $11.8\pm 1.5$ \\ % (47\%) \\
   % \hline
   C$_2$H$_4$ & - & $8.5\pm 0.8$ & $110\pm 7$ & $13.1\pm 1.2$ & $106\pm 7$ & $12.9\pm 1.3$ \\% (19\%)\\
   % \hline
   C$_2$H$_6$ & $100\pm 5$\cite{Carroll2002jcp} & $8.8\pm 0.8$ & - & - & $106\pm 7$ & $11.6\pm 0.8$\\ % (9\%)$^*$\\ 
   % \hline
   CH$_3$F & - & - & - & - & $104\pm 7$ & $11.5\pm 1.3$ \\ % (22\%) \\
   \hline
 \end{tabular}
 \caption{\label{tab:hydroC_dAD}
 Total AD widths and dAD branching ratios for C $1s^{-1}$ core hole states in small hydrocarbons.  For C$_2$H$_n$, an average over gerade and ungerade states is given.  All experimental values for \dadbr\ are taken from Ref.~\onlinecite{Roos2018scirep}.
 }
\end{table*}

In this subsection, we move to a set of small hydrocarbons to study the effect of multiple bonds and fluorine substitution. Calculated total decay widths and dAD branching ratios for the C $1s^{-1}$ states are compared to available experimental data in Tab.\ \ref{tab:hydroC_dAD}. Compared to Sec.\ \ref{sec:isoNe}, we have employed a smaller one-particle basis set to allow for a consistent description of all molecules. Still, even using the reduced basis, the full \adcf\ scheme was unfeasible for the largest molecules; thus, only Fano-\adcm\ results are available for C$_2$H$_6$ and CH$_3$F. The smaller basis set is behind the difference between the results given for methane in the present and the previous section. However, the lower quality of the representation of the continuum translates only into somewhat inferior stability of the Stieltjes imaging at higher orders, which is reflected by the larger error margins shown in Tab.\ \ref{tab:hydroC_dAD} (see SM for the comparison of Stieltjes imaging stability). The decay widths and branching ratios are still well converged.

Focusing first on the total AD widths, we observe again that Fano-ADC(2,2) yields slightly higher values than the experiment, with both tested schemes producing mutually consistent results. In agreement with the experimental data, Fano-ADC(2,2) predicts the largest width for C$_2$H$_2$ but overestimates the decay width for methane relative to other molecules. However, considering the sub-10\% differences between the molecules, we can only conclude that the calculations are consistent with the measurements within the error margins but are not accurate enough to reliably reproduce any variances in the AD widths across the sample.

Note also that for methane, the reported experimental values span a wide range from 83\,meV to 120\,meV \cite{Kolorenc2011jcp}. Crucially, the observed line width also depends on the energy of the ionizing radiation \cite{Carroll1999pra}. The effect was attributed to a different impact of post-collision interaction on the line shape used to determine the lifetime broadening. However, it might also be caused by an admixture of core-ionization satellites in the metastable state populated by the photoionization. Indeed, our calculations show that such satellite states become accessible for photon energies above 305\,eV, and their AD width is around $70\,\meV$. Even a low admixture can thus explain the observed decrease of the decay width with the photon energy.

Turning to the \dadbr, the experiment shows a 30\% rise going from CH$_4$ to C$_2$H$_2$ and then only minimal growth with an increasing number of hydrogens. Fano-ADC provides a qualitatively correct picture, capturing the difference between methane and larger molecules, but the rise of \dadbr\ is higher (50\%-60\%). Quantitatively, the \dadbr s are overestimated by 25\%-45\%, more than for the Ne-isoelectronic molecules. There is also no systematic trend reproduced among the C$_2$H$_n$ molecules. But again, the experimentally observed differences are well below the accuracy of the method, as well as the error margins of the experiment.

Finally, our results on fluoromethane confirm that halogen substitution leads to a decrease in the AD width, in agreement with the available literature.\cite{Carroll2002jcp,Zahl2012jelspec} In the CH$_{4-n}$Cl$_n$ series, the observed AD width decreases by about 9-10\% with each substituted chlorine.\cite{Zahl2012jelspec} For the fluorine substitution, our calculations predict a decrease of 4-5\%. The weaker effect can be attributed to the lower electron affinity of fluorine, as the reduction of AD rate is due to the depopulation of the carbon atomic orbitals.\cite{Walsh1994jpb}

The predicted \dadbr, on the other hand, increases by nearly 50\% compared to CH$_4$ and is comparable to C$_2$H$_n$. The available experimental data are consistent with such a trend. Roos \emph{et al.}\cite{Roos2018scirep} report \dadbr\ 16.5\% for CH$_3$Cl and 15.8\% for CF$_4$, supporting the supposition that \dadbr\ depends on the total number of valence electrons rather than just the electrons localized on the ionized atom, cf Eq.\ \eqref{eq:dADBR_Roos}. Unfortunately, both experimentally studied molecules are too large to be manageable by the present implementation of Fano-ADC(2,2).

\subsection{Discussion and in-depth analysis}
\label{sec:analysis}
\begin{figure}[htpb]
    \centering
    \includegraphics[width=0.95\linewidth]{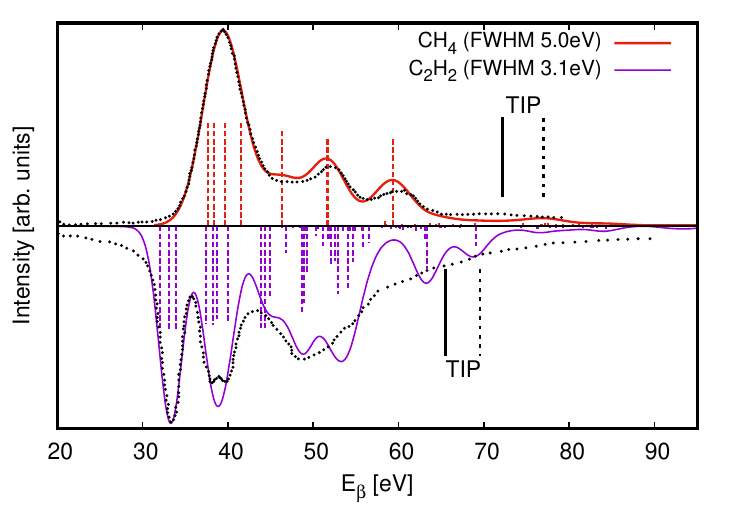}
    \caption{\label{fig:CH4_C2H2_Auger}Auger spectra of CH$_4$ and C$_2$H$_2$ as functions of the decay channel energy $E_\beta$.
    %\adcf, large basis.
    Experimental spectra shown by black dots are taken from Refs.\ \onlinecite{Kivimaki1996jpb} and \onlinecite{Kivimaki1997jpb} for CH$_4$ and C$_2$H$_2$, respectively. The experimental data are shifted by 2.8\,eV for CH$_4$ and 1.8\,eV for C$_2$H$_2$ towards lower $E_\beta$ to match the calculated position for the lowest, most intense peak. The coloured dashed sticks indicate positions and $2h$ strengths of the dicationic channels. Black vertical lines indicate the triple ionization potential [full -- ADC(2)x, dashed -- ADC(1)].
    % {\color{red} Add 1st and 2nd order TIPS.}
    %(average of ADC(1) and ADC(2)x values).
    %{\color{red}\emph{(i.e., the theory gives faster Auger electrons)}}.
    }
\end{figure}
\begin{figure}[htpb]
    \centering
    \includegraphics[width=0.95\linewidth]{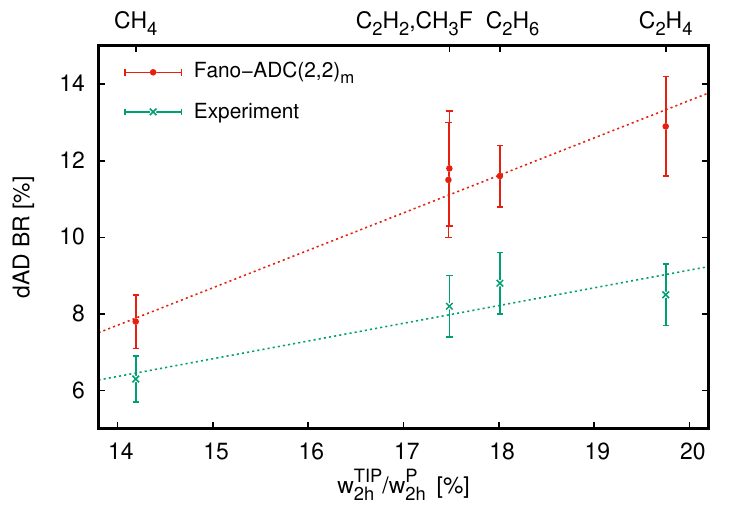}
    \caption{\label{fig:aboveTIP_CnHm}\dadbr\ as a function of the $2h$ weight cumulated in dicationic states above TIP.}
\end{figure}

In the previous two sections, we presented calculated \dadbr s for a number of small molecules. The Fano-ADC(2,2) method captures all the qualitative trends and generally provides very good quantitative accuracy, but we have also observed somewhat non-uniform performance. For some systems, Ne in particular, it tends to underestimate the double decay while overestimating \dadbr\ for hydrocarbons. Notably, the higher the \dadbr, the closer the \adcm\ results to the \adcf\ scheme. In this section, we trace this behaviour to the inconsistency of the representation of the $3h2p$-like final states relative to $1h$- and $2h1p$-like ones and of the couplings between these classes.

To this end, let us study the C\,$1s$ AD spectrum in methane and ethyne shown in Fig.~\ref{fig:CH4_C2H2_Auger}. The spectra calculated using the Fano-\adcf\ method are convoluted with a Gaussian function of appropriate width to compare to the spectra measured by Kivim\"aki \emph{et al.}\cite{Kivimaki1996jpb,Kivimaki1997jpb} The experimental spectra are shifted towards lower double ionization energy $E_\beta$ and scaled to match the calculated position and intensity of the strongest peak. We shift the experimental rather than theoretical signal because the relative position of calculated double and triple ionization potentials is vital for the following discussion.

We observe that the Fano-\adcf\ method yields AD spectra in excellent agreement with the experiment, confirming the high accuracy of the channel-specific branching ratios. As detailed in Sec.~\ref{sec:BRs}, the population of the dicationic channels above TIP is used to determine the \dadbr, together with the intensity associated with the complementary projector $\hat{P}_\perp$. Focusing on the CH$_4$ spectrum, there is a small bump just above TIP, associated with almost purely satellite $3h1p$-like dicationic final states. The energies of such states are calculated through first-order PT only and are, thus, less accurate than those of the main states and typically overestimated. Indeed, there is a hint of a similar bump at lower energy below TIP also in the experimental spectrum. Subtracting its contribution would reduce the calculated \dadbr\ to about 7\%, already in excellent agreement with the measured value $6.3\pm 0.6\%$. In ethyne, the AD signal associated with DI satellites around the TIP is even stronger, which is likely one of the reasons for the still larger overestimation of the respective \dadbr. 

The population of the satellite DI states by dAD is determined by the strength of the electronic correlation in the molecule, which is reflected by the mixing of $2h$ and $3h1p$ ISs in the correlated dicationic channels. In particular, we found that the calculated \dadbr\ is, to a large extent, determined by $w_{2h}^\mathrm{TIP}$, the cumulated contribution of $2h$ ISs in ADC(2)x dicationic channels above TIP. In Fig.\ \ref{fig:aboveTIP_CnHm}, we plot the dependence of both calculated and measured \dadbr s in hydrocarbons as a function of $w_{2h}^\mathrm{TIP}/w_{2h}^P$, i.e., the above-TIP percentage of the total $2h$ strength in the $\Psub$ subspace. Both data sets show distinct linear dependences, but the calculated \dadbr\ grows almost twice as fast as the measured one.

This observation shows that the strength of the coupling of the $1h$-like core-hole state to the $2h1p$ and $3h2p$ ISs is not evaluated consistently in the Fano-ADC(2,2) method. This conclusion is further supported by the fact that the most undervalued \dadbr\ is in neon, where the dAD is determined purely by direct coupling to $3h2p$ double-continuum final states. Notably, in such a scenario, the \adcm\ scheme performs significantly worse than \adcf.

The exact origin of this inconsistency is difficult to track down. The neglected second-order\cite{Kolorenc2020jcp} direct $1h/3h2p$ coupling can contribute. This problem is common to both \adcf\ and \adcm\ schemes. The slightly inferior performance of \adcm\ suggests that the inconsistent level of description of $2h1p$- and $3h2p$-like states is also a significant factor. While the second-order $1h/3h2p$ couplings could, in principle, be added without a prohibitive increase in the computational cost, the latter problem is inherent to the method.

\section{Conclusions}
\label{sec:conclusions}

We presented the application of the Fano-ADC(2,2) method to computing total and partial Auger decay widths of molecular core-hole states, including explicit evaluation of double Auger decay branching ratios. The branching ratios are calculated using channel projectors constructed from correlated ADC(2)x dicationic states, in accord with the ADC(2,2) level of representation of the continuum-like final states.

The typical relative error of the total AD widths is around 10\%, with an apparent bias towards overestimating the experimental values. The calculated Auger decay spectra show excellent agreement with the experiment, demonstrating the high accuracy of partial decay widths. These results confirm that, for normal AD, the Fano-ADC(2,2) is on par or better than other available state-of-the-art theoretical methods.\cite{Matz2023jpchA, Inhester2012jcp, Skomorowski2021jcpB, Tenorio2022jctc, Gerlach2022pccp} %, Ghosh2017chp}

For the investigated sample of molecules, the average error of \dadbr\ compared to available experimental data is about 30\%, which we evaluate as an excellent result considering the complexity of the double decay process and the universality of the method. In contrast to the total AD widths, the \dadbr\ can be both over- and underestimated depending on the character of the dicationic states in the relevant spectral range.

One of the sources of the error is the uncertainty of the effective position of the triple ionization threshold, stemming from the different quality of the representation of the main and satellite final states. We have also identified a certain degree of disproportion between the magnitude of the couplings of the initial core-hole state and the two classes of final states. The inconsistency is worse in the computationally cheaper \adcm\ scheme, which is to be attributed to the even lower level of correlation in the $3h2p$ class.

It should be stressed that vibrational motion can significantly affect the decay rates of core hole states in molecules or even quench the second step of the cascade dAD by dissociating or closing the decay channels. The present study does not include any effects of nuclear dynamics or the vibrational wave packet broadening. These phenomena will be examined in a follow-up publication.

In summary, Fano-ADC(2,2) represents a practical tool to investigate molecular double Auger decay. Beyond the vibrational motion, future development should focus on the discrimination between simultaneous and cascade double decay. One possible strategy is the restricted active space ADC technique. Alternatively, the Fano-ADC(2)x methodology for dicationic states\cite{Kolorenc2008jcp} can be used to define purely bound-like dicationic channels in a reduced configuration subspace, i.e., repeating similar $\Qspace/\Pspace$ projection strategy as for the cationic configuration space. This approach will naturally produce also total widths and branching ratios for the intermediate metastable dicationic states, offering a complete description of the cascade double Auger decay.

% \begin{itemize}
%     \item It should be noted that owing to the unfavourable computational scaling of the ADC(2,2) scheme, even using significantly smaller one-particle basis sets does not automatically grant access to significantly larger molecules.
%     \item the error in \adcm\ \dadbr\ is not universal; it depends on the character of the DI states above TIP
%     \item one of the possible conclusion could be that additional 20\% uncertainty need to be added to \adcm\ dAD BRs
%     \item It seems that the coupling to $3h2p$ ISs is underestimated relative to the coupling to $2h1p$ couplings
%     \begin{itemize}
%         \item \emph{could this be just due to the neglected 2nd order corrections to the $N-1$-to-$N-2$ projectors?}
%         \item \emph{the spectra are very good, but we are talking about few percent overal...}
%     \end{itemize}
%     \item vibrations are not included in any way
%     {\color{red}It should be stressed that any nuclear dynamics or just the spread of the vibrational wave packet is neglected in the present study, which might quench the second step of the cascade dAD by dissociation or closing the channel.}
%     \item should we comment on Auger spectra rel. to Jagau? e.g., automatic explicit inclusion of satellite states...?
% \end{itemize}

\section*{Supplementary material}
See supplementary material for detailed information about implementation, basis sets and other computational details.

\begin{acknowledgments}
Financial support by the Czech Science Foundation (Project GA\v{C}R No.\ 22-22658S) is gratefully acknowledged.
\end{acknowledgments}

\section*{Data availability}
The data supporting this study's findings are available from the corresponding author upon reasonable request.

\bibliography{papers}

\appendix

\section{Supplementary material}

\subsection{Implementation}

In the present implementation, we use the MOLCAS\cite{molcas74} quantum chemistry package for HF calculations and evaluation of the Coulomb two-electron integrals. Alternatively, the two-electron integrals can be computed using the GBTOlib library\cite{Masin2020cpcom}, which allows augmenting the multi-centre GTO basis by a continuum-like B-spline type orbitals (BTO). The latter option was not used in the present work and will be covered in a future publication on Fano-RCS-ADC. The calculations of AD widths, secondary electron spectra and dAD branching ratios for a given molecule proceed as follows:
\begin{enumerate}
    \item Optimal GTO basis is chosen based on the convergence of the total and partial AD widths and stability of the Stieltjes imaging procedure using the computationally cheap Fano-ADC(2)x scheme.
    \item Extended second-order TI-ADC(2)x calculations are performed to determine the triple ionization potential (TIP) and the number of open tri-cationic decay channels. First-order TIP$^{(1)}$ is also computed.
    \item DI-ADC(2)x calculations are performed to obtain the channel energies $E_\beta^{(N-2)}$ and channel wave functions $\ket{\varphi_\beta^{(N-2)}}$. Full diagonalization of the Hamiltonian matrix was possible in all cases. Alternatively, accurate results can be obtained using a filter diagonalization method\cite{Mandelshtam1997jcp,Mandelshtam2003jtcc} (FDM) focused on the relevant lower part of the DI spectrum.
    \item ADC(2,2) IS space is generated and, based on the results of steps 2 and 3, decomposed into the $\Qspace$ and $\Pspace$ subspaces using the general \emph{scheme B}, see Ref.~\onlinecite{Kolorenc2020jcp}.
    \item $\mathbf{QMQ}$ Hamiltonian matrix is diagonalized using the FDM method in the spectral region of the inner-shell vacancy of interest, and the appropriate discrete state $\ket{\phi_d}$ is selected among the eigenvectors based on the overlap with the relevant core-vacancy configuration.
    %We note that the convergence toward $\ket{\phi_d}$ is very fast and robust if the generation of the Krylov basis starts from the appropriate eigenvector of the comparatively small sub-block of $\mathbf{QMQ}$ corresponding to the $1h$ and $2h1p$ subspace.
    \item $\ket{\theta}=\mathbf{PM}\ket{\phi_d}$ and $\ket{\theta_\beta}=\mathbf{P_\beta^{(N-2)} M}\ket{\phi_d}\ (\beta=1,\dots,N_c$) projections are evaluated. % using the spin-free form of the formulas given in Appendix \ref{app:DIprojectors}.
    \item %\label{step:symortho}Symmetric orthogonalization $\ket{\theta_\beta}\xrightarrow{S_\theta}\ket{\tilde{\theta}_\beta}$ is performed and the 
    The complementary projection is constructed as $\ket{\theta^\perp}=\mathbf{P^\perp M}\ket{\phi_d}=\ket{\theta}-\sum_\beta \ket{{\theta}_\beta}$.
    \item The so-called Lanczos pseudospectrum of the $\mathbf{PMP}$ Hamiltonian matrix is obtained using block-Lanczos algorithm,\cite{Parlett1998} with the vectors $\ket{\theta}$, $\ket{{\theta}_\beta}$ and $\ket{\theta^\perp}$ defining the starting block.  The pseudospectrum can be readily used for the Stieltjes imaging procedure in place of the full set of exact eigenvectors of $\mathbf{PHP}$.\cite{Kopelke2011jcpB} It has been shown that converged spectral moments are obtained using a relatively low number of Lanczos iterations, much smaller than the matrix dimension\cite{Gokhberg2009jcp}. Furthermore, by this construction, the desired matrix elements $\bra{\chi_i}\mathbf{P}_{{\beta}}\mathbf{M}\ket{\phi_d}$ are obtained directly as elements of the so-called short eigenvectors in terms of the Krylov basis. Computationally expensive expansion of the long eigenvectors in terms of the full IS basis is thus avoided.
    %\item The matrix elements $\bra{\chi_i}\mathbf{P}_{\tilde{\beta}}\mathbf{M}\ket{\phi_d}$ are transformed back the original set of channel projectors $\hat{P}_\beta^{(N-2)}$ using the transformation matrix $S_\theta$ obtained in step \ref{step:symortho}.
    \item The total decay width $\Gamma$, partial decay widths $\Gamma_\beta$ and the dAD branching ratio \eqref{eq:dAD BR} are computed using the Stieltjes imaging procedure.
    \item Auger electron spectrum $I_{AD}(E)$ can be modelled as a sum of Lorentzian profiles of the width $\Gamma$ with positions defined by $E_d-E_\beta^{(N-2)}$ and intensities by $\Gamma_\beta$.
\end{enumerate}
All the ADC calculations are performed exploiting the largest Abelian subgroup of the molecular symmetry group. The computational cost is further reduced by using spin-free working equations, leading to a decoupled block structure of the matrix representations of various operators. In particular, only [$S=1/2$, $m=1/2$] projection of the $(N-1)$-electron system is considered, where $S$ and $m$ are the total spin and its projection, respectively. DI-ADC(2)x calculations are then performed independently for $S=0$ and [$S=1$, $m=0$] spin states. Owing to the independence of the non-relativistic Hamiltonian on $m$, the same expansion vector $\mathbf{q}_\beta$ [cf Eq.~\eqref{eq:channel_functions}] can be used for all values of $m$ when evaluating projectors for triplet dicationic channels.

\subsection{Stieltjes imaging}
\label{SMsec:SI}
The Stieltjes imaging procedure\cite{Langhoff1979empmc,Hazi1979empmc,Reinhardt1979cpcom} utilizes the negative spectral moments
\begin{equation}
    \label{eq:spectralSM}
    S_{-k}=\int\! E^{-k}\Gamma(E)dE \approx \sum_i (\epsilon_i)^{-k} \gamma_i.
\end{equation}
to recover the decay width function $\Gamma(E)$. At each order $n_S$, the lowest $2 n_S$ moments are used to construct an $n_S$-point integration rule (Gaussian quadrature) with the unknown $\Gamma(E)$ as the weight function,
\begin{equation}
    \int\! \Gamma(E) f(E) dE \approx \sum_{i=1}^{n_S} w_i^{n_S} f(E_i^{n_S}).
\end{equation}
Here, $f(E)$ is a smooth test function. This quadrature provides a piece-wise approximation of the correctly normalized cumulative distribution,
\begin{equation}
    F(E) = \int_{E_\mathrm{min}}^E \Gamma(E') dE' \approx \sum_{i=1}^q w_i^{n_S},\quad E_q^{n_S}<E\le E_{q+1}^{n_S}.
\end{equation}
Finally, the discrete $n_S-1$-point sampling of the decay width function is obtained through the Stieltjes derivative as
\begin{equation}
    \Gamma^{n_S}(\bar{E}) = \frac{1}{2}\frac{w_{q+1}^{n_S}+w_q^{n_S}}{E_{q+1}^{n_S}-E_q^{n_S}}
\end{equation}
with $\bar{E}=\frac{1}{2}(E_{q+1}^{n_S}+E_q^{n_S})$.

\begin{figure}[ht]
    \centering
    \includegraphics[width=0.95\linewidth]{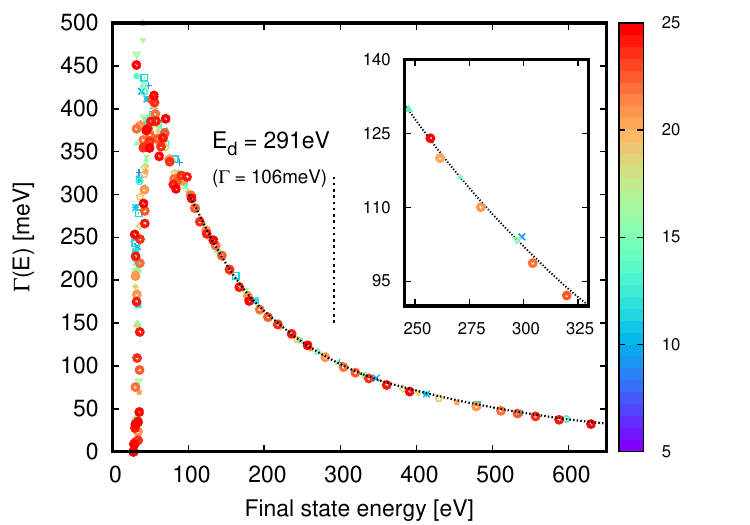}
    \caption{\label{fig:SI.CH4.large}Output of the Stieltjes imaging procedure -- CH$_4$, Fano-\adcf, large basis set (cf Tab \ref{tab:isoNe_Gamma}). Points of different colours show the raw output of the Stieltjes imaging procedure at orders $n_S=5-25$. The dotted line is the interpolation through orders $n_S=10-14$. The inset shows the region around the discrete state energy $E_d$.}
\end{figure}

\begin{figure}[ht]
    \centering
    \includegraphics[width=0.95\linewidth]{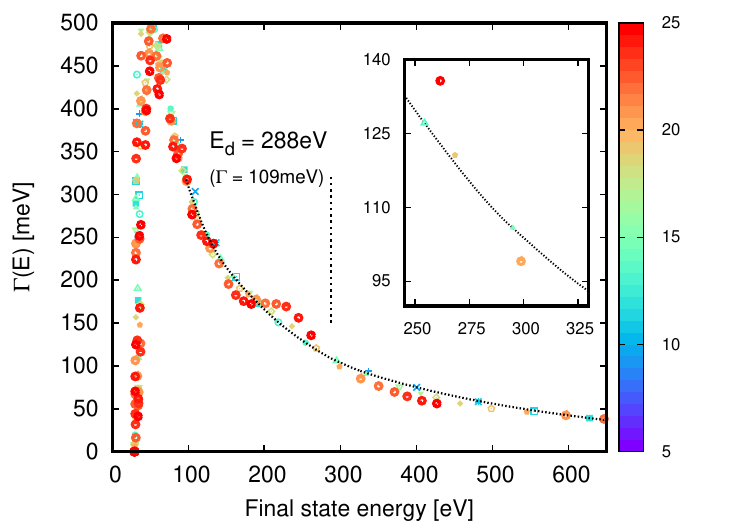}
    \caption{\label{fig:SI.CH4.small}Output of the Stieltjes imaging procedure -- CH$_4$, Fano-\adcm, small basis set (cf Tab.\ \ref{tab:hydroC_dAD}). Points of different colours show the raw output of the Stieltjes imaging procedure at orders $n_S=5-25$, and the dotted line is the interpolation through orders $n_S=9-12$.}
\end{figure}

This procedure is performed for a series of orders $n_S$. Finally, monotonicity-preserving interpolation of points obtained at several consecutive orders in the region of convergence yields the desired approximation to $\Gamma(E)$ and, in particular, $\Gamma(E_d)$. The stability of the algorithm depends on the quality of the input spectral moments, which is, in turn, determined by the quality of the one-particle basis. Two examples are shown in Figs.\ \ref{fig:SI.CH4.large} and \ref{fig:SI.CH4.small} for the CH$_4$ molecule. In the former, computed using the Fano-\adcf\ method with the large basis set, we observe that the procedure is very stable up to the highest orders, at least above 100\,eV final state energy. In the latter, obtained using the Fano-\adcm\ scheme with a comparatively small basis set, the algorithm gradually loses stability above $n_S\approx 17$. Still, a sufficiently accurate result is obtained at lower orders in both cases.

\afterpage{\clearpage}

\subsection{Basis sets and other computational details}
% \label{app:Bases}
%
Here, we provide information on the basis sets and other details of calculations reported in the present work. All standard basis sets were downloaded from the basis set exchange library \cite{pritchard2019a} or taken directly from the library supplied with MOLCAS\cite{molcas74}.

\subsubsection{Sec.\ \ref{sec:isoNe}, Tabs.\ \ref{tab:isoNe_Gamma}, \ref{tab:isoNe_dAD}}
\begin{itemize}
    \item Ne \\
    \underline{primitive basis:} uncontracted aug-cc-pV6Z (up to H-type functions) augmented by 19s, 18p, 15d, 10f and 5g even-tempered exponents. The exponents of the augmentation are given in Tab.\ \ref{tab:even_tempered_aug}. \\
    \underline{basis and configuration space restrictions:} The one-particle basis is restricted to orbital energies up to 800\,Ha, resulting in 396 active HF orbitals.

    \item HF \\
    \underline{primitive basis:} uncontracted aug-cc-pCV5Z (up to G-type functions) basis augmented by 7s,5p,4d,3f even-tempered exponents on F, uncontracted aug-cc-pVQZ basis augmented by 3s, 3p even-tempered functions on H. The exponents of the even-tempered augmentations are given in Tab.\ \ref{tab:even_tempered_aug}. Additional 10s, 10p, 8d, 5f continuum-like Kaufmann-Baumeister-Jungen continuum-like functions\cite{Kaufmann1989jpb} (KBJ) placed at the centre of mass, see Tab.\ \ref{tab:CM_basis}. \\
    \underline{basis and configuration space restrictions:} The one-particle basis is restricted to orbital energies up to 800\,Ha, resulting in 365 active HF orbitals. In addition, the $3h2p$ class is restricted to 60\,Ha zero-order energy and a maximum of one F$(1s)$ core hole.

    \item H$_2$O \\
    \underline{primitive basis:} uncontracted aug-cc-pCV5Z (up to G-type functions) basis with no additional augmentation on O, uncontracted d-aug-cc-pVQZ basis with no additional augmentation on H. Additional 10s, 10p, 8d, 5f continuum-like KBJ continuum-like functions placed at the centre of mass with the same exponents as for HF, cf Tab.\ \ref{tab:CM_basis}.\\
    \underline{basis and configuration space restrictions:} The one-particle basis is restricted to orbital energies up to 200\,Ha, resulting in 436 active HF orbitals.

    \item NH$_3$ \\
    \underline{primitive basis:} uncontracted aug-cc-pCV5Z (up to G-type functions) basis further augmented by 6s, 5p, 4d even-tempered exponents on N (the exponents of the augmentation are given in Tab.\ \ref{tab:even_tempered_aug}), aug-cc-pVQZ basis with no additional augmentation on H.
     Additional 10s, 10p, 8d, 5f continuum-like KBJ continuum-like functions placed at the centre of mass with the same exponents as for HF, cf Tab.\ \ref{tab:CM_basis}.
    \\
    \underline{basis and configuration space restrictions:} The one-particle basis is restricted to orbital energies up to 170\,Ha, resulting in 389 active HF orbitals. In addition, the $3h2p$ class is restricted to 30\,Ha zero-order energy.

    \item CH$_4$ \\
    \underline{primitive basis:} uncontracted aug-cc-pCV5Z (up to G-type functions) basis with no additional augmentation on C, aug-cc-pV5Z basis augmented by 5s, 5p, 3d continuum-like KBJ continuum-like functions on H (the exponents of the augmentation are given in Tab.\ \ref{tab:even_tempered_aug}).
     Additional 10s, 10p, 8d, 5f diffuse continuum-like KBJ continuum-like functions placed at the centre of mass, see Tab.\ \ref{tab:CM_basis}.
    \\
    \underline{basis and configuration space restrictions:} The one-particle basis is restricted to orbital energies up to 60\,Ha, resulting in 619 active HF orbitals. In addition, the $3h2p$ class is restricted to 20\,Ha zero-order energy.

\end{itemize}

    \begin{table}[ht]
        \centering
         {\tiny
        \begin{tabular}{r|r|r|r|r}
        \multicolumn{5}{c}{\textbf{Ne}} \\
        \multicolumn{1}{c|}{\textbf{s}} & \multicolumn{1}{c|}{\textbf{p}} & \multicolumn{1}{c|}{\textbf{d}} & \multicolumn{1}{c|}{\textbf{f}} & \multicolumn{1}{c}{\textbf{g}} \\
        \hline
        257.63655 &	110.00264 &	8.79082 &	6.85533 &	0.08516 \\
109.45154 &	38.44458 &	3.83090 &	3.00434 &	0.12599 \\
48.56314 &	15.18159 &	1.66942 &	1.31621 &	0.20512 \\
22.29541 &	6.46461 &	0.72730 &	0.55150 &	0.39131 \\
10.43110 &	2.86324 &	0.62256 &	0.16630 &	1.02186 \\
4.59778 &	1.27403 &	0.29960 &	0.10232 \\	
1.98686 &	0.55698 &	0.24216 &	0.06924 \\
0.88116 &	0.43008 &	0.12784 &	0.04995 \\	
0.37848 &	0.23509 &	0.07883 &	0.03773 \\	
0.09850 &	0.16934 &	0.05343 &	0.02950 \\	
0.05273 &	0.08989 &	0.03858 \\		
0.03277 &	0.03777 &	0.02916 \\	
0.02233 &	0.02731 &	0.02282 \\	
0.01618 &	0.02067 &	0.01833 \\	
0.01226 &	0.01618 &	0.01506 \\	
0.00961 &	0.01301 \\			
0.00774 &	0.01069 \\			
0.00636 &	0.00894 \\			
0.00532 \\				
       \hline
        \hline
        \multicolumn{5}{c}{\textbf{F}} \\
        \multicolumn{1}{c|}{\textbf{s}} & \multicolumn{1}{c|}{\textbf{p}} & \multicolumn{1}{c|}{\textbf{d}} & \multicolumn{1}{c|}{\textbf{f}} & \multicolumn{1}{c}{\textbf{g}} \\
        \hline
        152.27868 &	49.96212 &	36.39523 &	10.98067 \\
    79.08936 &	28.13557 &	12.98586 &	3.34919 \\
    65.47814 &	22.58535 &	4.85060 &	1.28934 \\
    54.20940 &	10.70677 &	1.89550 \\
    28.27740 &	4.16751 \\
    11.91002 \\			
    5.32181 \\
        \hline
        \hline
        \multicolumn{5}{c}{\textbf{N}} \\
        \multicolumn{1}{c|}{\textbf{s}} & \multicolumn{1}{c|}{\textbf{p}} & \multicolumn{1}{c|}{\textbf{d}} \\
        \hline
        40.34699 &	30.73722 &	23.43513 \\
    17.37361 &	17.29104 &	8.08111 \\
    7.26460 &	6.55639 &	2.90259 \\
    3.27353 &	2.51795 &	1.13216 \\
    1.38309 &	1.01823 \\
    0.55261 \\
\hline
\hline
\multicolumn{5}{c}{\textbf{H} (even-tempered for HF)} \\
\multicolumn{1}{c|}{\textbf{s}} & \multicolumn{1}{c|}{\textbf{p}} \\
        \hline
        5.91995 &	1.38589 \\
        1.50090 &	0.49467 \\
        0.45375 &	0.15736 \\
        \hline
        \multicolumn{5}{c}{\textbf{H} (KBJ for CH$_4$)} \\
\multicolumn{1}{c|}{\textbf{s}} & \multicolumn{1}{c|}{\textbf{p}} & \multicolumn{1}{c|}{\textbf{d}} \\
        \hline
        0.24565 &	0.43008 &	0.62256 \\
        0.09850 &	0.16934 &	0.24216 \\
        0.05273 &	0.08989 &	0.12784 \\
        0.03277 &	0.05561 \\
        0.01618 &	0.03777 \\
        \end{tabular}}
        \caption{Exponents of atom-centered augmentations of the standard basis sets used in Sec.\ \ref{sec:isoNe}, Tabs.\ \ref{tab:isoNe_Gamma}, \ref{tab:isoNe_dAD}.}
        \label{tab:even_tempered_aug}
    \end{table}

        \begin{table}[ht]
        \centering
         {\tiny
        \begin{tabular}{r|r|r|r}
\multicolumn{4}{c}{\textbf{HF, H$_2$O, NH$_3$}} \\
\multicolumn{1}{c|}{\textbf{s}} & \multicolumn{1}{c|}{\textbf{p}} & \multicolumn{1}{c|}{\textbf{d}} & \multicolumn{1}{c|}{\textbf{f}} \\
\hline
0.24565 &	0.43008 &	0.62256 &	0.82035 \\
0.09850 &	0.16934 &	0.24216 &	0.31628 \\
0.05273 &	0.08989 &	0.12784 &	0.16630 \\
0.03277 &	0.05561 &	0.07883 &	0.10232 \\
0.02233 &	0.03777 &	0.05343 &	0.06924 \\
0.01618 &	0.02731 &	0.03858 \\	
0.01226 &	0.02067 &	0.02916 \\	
0.00961 &	0.01618 &	0.02282 \\	
0.00774 &	0.01301 \\		
0.00636 &	0.01069 \\
\hline
\hline
\multicolumn{4}{c}{\textbf{CH$_4$}} \\
\multicolumn{1}{c|}{\textbf{s}} & \multicolumn{1}{c|}{\textbf{p}} & \multicolumn{1}{c|}{\textbf{d}} & \multicolumn{1}{c|}{\textbf{f}} \\
\hline
0.05273 &	0.08989 &	0.62256 &	0.16630 \\
0.03277 &	0.05561 &	0.24216 &	0.10232 \\
0.02233 &	0.03777 &	0.12784 &	0.06924 \\
0.01618 &	0.02067 &	0.05343 &	0.04995 \\
0.01226 &	0.01618 &	0.03858 &	0.03773 \\
0.00961 &	0.01301 &	0.02916 \\
0.00774 &	0.01069 &	0.02282 \\
0.00636 &	0.00894 &	0.01833 \\
0.00532 &	0.00758 \\
0.00452 &	0.00652
        \end{tabular}}
        \caption{Exponents of the continuum-like basis set placed at the centre of mass for calculations reported in Sec.\ \ref{sec:isoNe}, Tabs.\ \ref{tab:isoNe_Gamma}, \ref{tab:isoNe_dAD}.}
        \label{tab:CM_basis}
    \end{table}

\afterpage{\clearpage}

\subsubsection{Sec.\ \ref{sec:hydrocarb}, Tabs.\ \ref{tab:hydroC_dAD}}
\noindent
\underline{primitive basis:}
\begin{itemize}
    \item C, F: aug-cc-pCVQZ (up to F-type functions) with no additional augmentation
    \item H: aug-cc-pVQZ with no additional augmentation
\end{itemize}
The one-particle basis is restricted to orbital energies up to 100\,Ha for all molecules. The F $1s$ orbital was frozen (inactive) for the CH$_3$F calculations.

\noindent
\underline{configuration space restrictions:}
\begin{itemize}
    \item the $3h2p$ class restricted to 20\,Ha zero-order energy and a maximum of 1 C($1s$) hole
\end{itemize}

\afterpage{\clearpage}

% \section*{References}
% \bibliography{papers}

% \section{Trash}
% \input{trash}

\end{document}